\documentclass[12pt,preprint]{aastex}
\usepackage{multirow}

\shorttitle{Discovering bright quasars at intermediate redshifts}
\shortauthors{Wu et al.}

\begin{document}

\title{Discovering bright quasars at intermediate redshifts based on optical/near-IR colors}

\author{Xue-Bing Wu, Wenwen Zuo, Jinyi Yang, Qian Yang, Feige Wang\\Department of Astronomy,  Peking University, Beijing 100871, China\\ email: wuxb@pku.edu.cn}
\begin{abstract}
Identifications of quasars at intermediate redshifts ($2.2<z<3.5$) are  inefficient in most previous quasar surveys as their optical colors are similar to 
those of stars. The near-IR K-band excess technique has been suggested to overcome this difficulty. Our recent study also proposed to use optical/near-IR colors for selecting $z<4$ quasars. To verify the effectiveness of this method, we selected a list of 105 unidentified bright targets with $i\leq18.5$ from the quasar candidates of SDSS DR6 with both SDSS ugriz optical and UKIDSS YJHK near-IR photometric data, which satisfy our proposed Y-K/g-z criterion and have photometric redshifts between 2.2 and 3.5 estimated from the 9-band SDSS-UKIDSS data. 43 of them were observed with the BFOSC instrument on the 2.16m optical telescope at Xinglong station of NAOC in the spring of 2012. 36 of them were spectroscopically identified as quasars with redshifts between 2.1 and 3.4.
High success rate of discovering these quasars in the SDSS spectroscopic surveyed area further
demonstrates the robustness of both the Y-K/g-z selection criterion and the photometric redshift estimation technique. We also used the above criterion to investigate the possible star contamination rate to the quasar candidates of SDSS DR6, and found that it is much higher in selecting $3<z<3.5$ quasar candidates than selecting lower redshift ones ($z<2.2$).  The significant improvement in the photometric redshift estimation by using the 9-band SDSS-UKIDSS data than using the 5-band SDSS data is demonstrated and  a catalog of 7,727 unidentified quasar candidates in SDSS DR6 selected with the optical/near-IR colors and with photometric redshifts between 2.2 and 3.5 is provided. We also tested the Y-K/g-z selection criterion with the recently released SDSS-III/DR9 quasar catalog, and found 96.2\% of 17,999 DR9 quasars with UKIDSS Y and K-band data satisfy our criterion. With some available samples of red quasars and type II quasars, we find that 88\%  and 96.5\% of them can be selected by the Y-K/g-z criterion respectively, which supports that using the Y-K/g-z criterion we can efficiently select both unobscured and obscured quasars.
We discuss the implications of our results to the ongoing and upcoming large 
optical and near-IR sky surveys.
\end{abstract}
\keywords{galaxies: active --- galaxies:high-redshift --- quasars: general 
--- quasars: emission lines }

\section{Introduction}
Quasars are important 
extragalactic objects in astrophysics due to their unusual properties. They not only can 
be used to probe the physics of supermassive black holes and accretion/jet 
process, but also are closely related to the studies of galaxy 
evolution, intergalactic medium, large scale structure and cosmology. The number of quasars has increased steadily since their discovery in 1963 (Schmidt 1963; Hazard, Mackey \& Shimmins 1963; Oke 1963; Greenstein \& Matthews 1963).  Especially, a large number of quasars have been discovered in the last two decades in two large spectroscopic surveys, namely, the Two-Degree Fields (2DF) quasar survey (2QZ; Boyle et al. 2000) and the Sloan Digital Sky Survey (SDSS; York et al. 2000). 2DF mainly selected low redshift ($z < 2.2$) quasar candidates with UV-excess (Smith et al. 2005) and has discovered more than 20,000 blue quasars (Croom et al. 2004), while SDSS adopted a multi-band optical color selection method 
(Richards et al. 2002) and has identified more than 120,000 quasars (Schneider et al. 2010). 90\% of SDSS quasars have low redshifts ($z < 2.2$), though some dedicated methods were also proposed for finding high redshift quasars ($z > 3.5$) (Fan et al. 2001a,b; Richards et al. 2002). 

However, in the redshift range $2.2<z<3.5$, the selection of SDSS quasars is inefficient. Richards et al. (2006)
have demonstrated this problem by checking the efficiency of SDSS quasar selection with the FIRST radio quasars and found that the efficiency drops substantially in the redshift range $2.2<z<3.5$. There are two remarkable dips, one around $z=2.8$ and another around $z=3.4$.
The reason for this problem is well understood.
At redshift $2.2<z<3.5$,  the 
spectral energy distributions of quasars show similar optical
colors to that of normal stars, and quasar selections using the optical 
color-color diagrams 
become very inefficient due to the serious contaminations
of stars (Fan 1999; Richards et al. 2002; 2006; Schneider et al. 2007; Hennawi et al. 2010). 
In addition, SDSS preferentially selects $3<z<3.5$ quasars with intervening H I Lyman limit systems, which
can also result in a lower efficiency in identifying quasars with redshift around $z=3.4$ (Worseck \& Prochaska 2011).
Because of the 
importance of using Ly$\alpha$ forest of
$z>2.2$ quasars to study   cosmic baryon acoustic 
oscillation
(BAO)  (White 2003; McDonald \& Eisenstein 2007) and  using these quasars to construct the accurate 
luminosity function to study  
quasar evolution in the mid-redshift universe (Wolf et al. 2003; Jiang et al. 2006),  
we need to explore other more efficient ways  to 
identify the $2.2<z<3.5$ quasars than using  optical colors alone. We notice that significant efforts have been made
recently for the quasar target selections in the  SDSS-III Baryon Oscillation 
Spectroscopic Survey (BOSS; Eisenstein et al. 
2011, Ross et al. 2012), and much more quasars at intermediate redshifts have been found in SDSS-III/DR 9
(Paris et al. 2012) .

 One possible way to identify the $2.2<z<3.5$ quasars is
to use optical variability, as this is one of the well known quasar properties
(Hook et al. 1994; Cristiani et al. 1996; Giveon et al. 1999). Selecting quasars based on
variability is usually thought to be less biased to the optical colors, though more variation in the
shorter wavelength have been found for SDSS quasars (Vanden Berk et al. 2004; Zuo et al. 2012).
Recently, Schmidt et al. (2010), MacLeod et al. (2011) and  Butler \& Bloom (2011)  have proposed to select quasar candidates by constructing
various intrinsic
variability parameters from 
the light-curves of known quasars in SDSS Stripe 82 (hereafter S82; see also Sesar et al. 2007).
They claimed  
that with their methods they can efficiently separate quasars from stars and 
substantially  increase  the number of quasars at
 $\rm 2.5<z<3.0$. Moreover,  recent results from SDSS-III/BOSS (Eisenstein et al. 
2011) also
confirmed the high success rate of spectroscopically identifying variability 
selected quasars, which leads to a significant increase 
of $z>2.2$ quasar density  in S82 than that based on optical color selected quasars
(Palanque-Delabrouille et al. 2011; Ross et al. 2012). However, only for very limited sky areas 
the multi-epoch observational data have been publicly available, so at present the variability methods can not be  broadly used
for selecting quasars over  a large sky area.

Another possible way for separating $z>2.2$ quasars from stars is to utilize
their near-IR colors. Due to the different radiative mechanisms of stars and quasars, the continuum emission from
stars usually has a blackbody-like spectrum and decreases more rapidly from optical to near-IR wavelengths than
that of quasars, which usually display a power-law spectra over a broad range of wavelength plus thermal emissions from accretion disks and dusts. This leads to obvious color differences in near-IR band between stars and quasars, even though their optical spectra are similar. Because of this difference,
 a K-band excess technique has been proposed for identifying quasars at $z>2.2$ 
(e.g. Warren, Hewett \& Foltz 2000; Croom, Warren \& Glazebrook 2001; Sharp et al. 2002; 
Hewett et al. 2006; Chiu et al. 2007;
Maddox et al. 2008,2012;  Smail et al. 2008; Wu \& Jia 2010).  
Based on a
sample of 8498 quasars and a sample of 8996  stars complied from
the photometric data in the
ugriz bands of SDSS  and YJHK bands of UKIRT InfraRed Deep Sky Surveys (UKIDSS\footnote{
The UKIDSS project is defined in Lawrence et al. (2007). UKIDSS uses the UKIRT Wide Field Camera (WFCAM; Casali et al. 2007) and a photometric system described in Hewett et al. (2006). The pipeline processing and science archive are described in Hambly et al. (2008).}), Wu \& Jia (2010)
 proposed an efficient empirical criterion, (i.e. $\rm Y-K>0.46(g-z)+0.82$, where YK magnitudes are Vega 
magnitudes and $\rm gz$ magnitudes are AB magnitudes) for selecting  $z<4$
 quasars.  A check with the VLA-FIRST 
(Becker, White \& Helfand 1995) radio-detected 
SDSS quasars, which are thought to be free of color selection bias (see McGreer, Helfand  \& White (2009), however), 
also proved that with this Y-K/g-z criterion they can achieve the 
completeness higher than 95\% for these radio-detected quasars at $z<3.5$, which seems 
to be difficult when using the SDSS optical color selection criteria 
alone where two dips around $z\sim 2.7$ and $z\sim 3.4$ obviously exist
(Richards et al. 2002,2006; Schneider et al. 2007,2010). 
Recently, Peth, Ross \& Schneider (2011) extended the
study of Wu \& Jia (2010) to a larger sample of 130,000 SDSS-UKIDSS selected
quasar candidates and re-examined the near-IR/optical colors of them.

Although by combining the variability and
optical/near-IR color we may achieve the
maximum efficiency in identifying $2.2<z<3.5$ quasars (Wu et al. 2011), for most sky areas we
still lack publicly available variability data. One may also think of using radio and X-ray data, but they can only be helpful
for selecting specific quasar samples(White et al. 2000; Green et al. 1995), which represent only a small fraction of the whole quasar population.
Therefore, using optical/near-IR colors is probably still the most important way for selecting  $2.2<z<3.5$ quasars. Although we have done some observations to identify a few $z>2.2$ quasars (Wu et al. 2010a,b; Wu et al. 2011), more efforts are still needed  to check whether using our Y-K/g-z criterion can help us to
discover more quasars at  $2.2<z<3.5$, especially in the SDSS spectroscopically surveyed area.

The paper is organized as follows. In Section 2 we will describe the target  
selections and spectroscopic observations, and report our discovery of 36 new $2.2<z<3.5$ quasars. In section 3 we will use our  Y-K/g-z criterion to check the possible contaminations of stars in the quasar candidate catalog of SDSS DR6 and  investigate the improvement in the photometric redshift estimation by using the 9-band SDSS-UKIDSS data.  In section 4 we will check the effectiveness of our  Y-K/g-z criterion in selecting quasars with the recently released SDSS-III/Dr 9 quasar catalog. In section 5 we will check whether we can use the Y-K/g-z criterion to select red quasars and type II quasars. In section 6 we will estimate how many $2.2<z<3.5$ quasars we can select in additional to the SDSS-III/BOSS quasars. Finally we will discuss our results and their implications
in Section 7.

\section{Target selections and spectroscopic observations}
\subsection{Target selections}
The main purpose of our study is to use our proposed Y-K/g-z criterion to
discover more quasars at  $2.2<z<3.5$ in the SDSS spectroscopically surveyed area.
We started from the 1 million quasar candidates in SDSS DR6 given by Richards et al. (2009) (hereafter R09) from the
Bayesian classification, and selected 8,845 unidentified quasar candidates brighter than $i=18.5$. From them
we further selected 2,149 candidates with  photometric redshifts given in R09 greater than 2.2 and  redshift probability higher
than 0.5. Then we cross-matched these quasar candidates with the UKIDSS/Large Area Survey(LAS) DR7 using a position offset within 3$''$ and got
401 candidates with both SDSS ugriz and UKIDSS YJHK photometric data. 126 of them satisfy
our $\rm Y-K>0.46(g-z)+0.82$ criterion for selecting $z<4$ quasars (Wu \& Jia 2010). 17 among
these 126 candidates have been spectroscopically identified as quasars  after the SDSS DR6, and 15 of them
have redshifts greater than 2.2. Another one candidate was identified as a white dwarf. After excluding these 18 known objects, we got a list of 108 quasar candidates.

Since these quasar candidates have 9-band SDSS and UKIDSS photometric data, the photometric redshifts
obtained from these 9-band data are more accurate than those given by the 5-band SDSS photometric data (Wu \& Jia 2010). We use our developed
program to estimate the photometric redshifts based on the 9-band data and found that 15 objects among 108 quasar candidates have photometric redshifts smaller than 2, while for other quasar candidates our results are consistent with those obtained in R09 from the SDSS photometric data. After excluding these 15 objects, we get a target list of 93 quasar candidates with reliable photometric redshifts between 2.2  and 3.6.  After the public release of UKIDSS/LAS DR8 in April, 2012, we also added other 12 quasar candidates, which have UKIDSS/LAS DR8 data and satisfy our Y-K/g-z selection criterion but were not included in the UKIDSS/LAS DR7, making our final list containing 105 quasar candidates. The procedures of our target selections are also listed in Table 1.

\subsection{Spectroscopic observations}
Among these 105 quasar candidates, 92 have Right Ascension (RA) between 7hr and 17hr.  In March to May of 2012, we made 8-night spectroscopic observations on 43  bright quasar candidates (sampling rate of 46.7\%) with photometric redshifts  at  $ 2.2<z_{ph}<3.5$ using the BAO Faint Object Spectrograph and Camera (BFOSC) of the 2.16m optical telescope at the Xinglong station of National Astronomical Observatories, Chinese Academy of Sciences (NAOC) .  A low resolution grism with the dispersion of 198\AA/mm,  the wavelength coverage from 3850 to 7000\AA,  and the spectral resolution of 2.97$\AA$ was used.  During our observations, the typical seeing varied from 1.5$''$ to 2.5$''$, so we adopted the long slit width of 1.8$''$ and 3.6$''$ accordingly (no choice in between). In Table 2 we summarize the parameters and details of the observations for these 43 quasar candidates, including the observation date,
exposure time, SDSS and UKIDSS magnitudes, photometric redshift, identification result and spectral redshift. 

Among the 43 quasar candidates, 36 were spectroscopically identified as quasars with redshifts from 2.1 to 3.4, 2 were identified as stars and 5 remain unidentified due to the
lower signal-to-noise ratios of their spectra. The spectra of 36 new quasars, after the standard flat-field corrections, flux and wavelength calibrations, were plotted in Fig. 1.

The spectra of these quasars were analyzed following the method described detailedly in Shen et al. (2008) and Shen \& Liu (2012). First the spectra were redshift corrected to the rest-frame and were corrected for the Galactic extinction using the extinction map of Schlegel et al. (1998). They are then fitted based on an IDL code MPFIT (Markwardt 2009). We fitted the spectra with the pseudo-continuum model consisting of the featureless non-stellar continuum and Fe II emissions. The featureless non-stellar continuum is assumed to be a power-law, so two free parameters (amplitude and slope) are required. Templates for Fe II emissions have been constructed from the spectrum of the narrow-line Seyfert 1 galaxy, I Zw I (Boroson \& Green 1992), by convolving with a velocity dispersion and shifting with a velocity. We used the UV template generated by Vestergaard \& Wilkes (2001) and Tsuzuki et al. (2006) in the wavelength range of 1000-3500\AA. 

After constructing the pseudo-continuum, the broad CIV component is fitted with two Gaussians and
the narrow component is fitted with one Gaussian. However, as the spectra of 13 quasars have low signal-to-noise ratio ($\sim$ 6.5), we used only one Gaussian to fit the whole CIV emission line profile for them. 
 We measured the Full Width at Half Maximum of CIV line (FWHM(CIV)), luminosity at 1350\AA ($L_{1350}$) from the spectra. We note that the line widths for  13 quasars with lower signal-to-noise ratio are rough estimates.  The black hole mass is calculated based on FWHM(CIV) and $L_{1350}$ with Eq.(7) in Vestergarrd \& Peterson (2006)(see also Kong et al. 2006). Using a scaling relation between $L_{1350}$ and bolometric luminosity $L_{bol}$, $L_{bol} = 4.62 L_{1350}$, we estimate the bolometric luminosity for these quasars. Based on the obtained black hole mass and bolometric luminosity, we also calculate their Eddington ratios($L_{\rm bol}/L_{Edd}$, where $L_{Edd}$ is the Eddington luminosity). The results are summarized in Table 3. Although we noticed that the uncertainties of these values are probably quite large due to the low spectral quality and the unusual properties of CIV, the overall properties of these quasars, including the line width, continuum luminosity, black hole mass and Eddington ratio, are consistent with those of typical SDSS quasars with redshift greater than 2.2 (Shen et al. 2011). The continuum slope parameter, $\alpha_{\lambda}$, is given for each quasar in Table 3. The median value is -1.315. If we convert it to $\alpha_{\nu}$, the median value of $\alpha_{\nu}$ is then -0.685, which is not too different from -0.517 and -0.862 obtained from SDSS DR9 and DR7 quasars, respectively (Paris et al. 2012).

We also investigate the Broad Absorption Line (BAL)  quasars in our sample. The Balincity Index is calculated for each quasar using the traditional method (Weymann et al. 1991).
14 quasars have positive BI values, indicating that they are probably BAL quasars. However, by visually inspecting the quasar spectra, we find that this traditional method risks identifying
false troughs from noisy and poor continuum fitting. To avoid these false identifications, we calculate the 
Balnicity Index (BI) and Absorption Index (AI), adding the same extra minimum depth and width requirement in the emission line region ( for more details see section 4.4 of Trump et al. (2006)). With these procedure we found 4 BALs. However, SDSS J124605.36+071128.2 is likely not a real BAL due to the lack of the spectrum shortward to the C IV line center. The remained 3 BALs are SDSS J115531.45-014611.9, 
SDSS J1359420+022426.0 and SDSS J142405.57+044105.5. Their BI values are 3379.27, 381.18 and 178.70 km/s, respectively.
 The high velocities (3837 - 21293 km/s) derived from their broad absorption features of the CIV lines are consistent with those of quasars with higher UV luminosity (Gibson et al. 2009). This is also expected if the BAL outflow is produced by the strong radiation pressure (Murray et al. 1995).

\subsection{Success rate of finding $2.2<z<3.5$ quasars}
Our spectroscopic observations identified 36 quasars  at $2.1<z<3.4$ and 2 stars from 43 candidates, which indicated a success rate of at least 83.7\% in identifying the bright quasars at intermediate redshifts because 5 candidates still remain unidentified. This high success rate is largely due to the quasar
candidate selection procedures we adopted, especially by using the  Y-K/g-z criterion and 9-band SDSS-UKIDSS photometric redshifts to 
select $2.2<z<3.5$ quasar candidates.  As we stated before,  using  photometric redshifts given in R09 greater than 2.2 and  redshift probability higher
than 0.5 enables us to reduce the number of quasar candidates brighter than $i=18.5$ from 8845 to 2149.
In addition, using the  Y-K/g-z criterion we can reduce the number of quasar candidates with SDSS-UKIDSS data from 401 to 126. Therefore, our high success rate of identifying $2.2<z<3.5$ quasar is not a surprise
because we can efficiently exclude the star contaminations by using the Y-K/g-z criterion and select most reliable
$2.2<z<3.5$  quasar candidates by using the photometric redshifts obtained from the SDSS or SDSS-UKIDSS photometric data.

\section{Accuracy of photometric redshifts and star contaminations in the quasar candidate catalog of SDSS DR6 }
We  selected the quasar targets from the SDSS DR6 1 million quasar candidate catalog of R09, and used both
the SDSS and UKIDSS photometric data for further selections and photometric redshift estimations to achieve
the high success rate of identifying $2.2<z<3.5$ quasars. 
With the SDSS-UKIDSS optical/near-IR data and our proposed quasar selection criterion, we may also investigate the accuracy of photometric redshifts and the possible star contaminations in the quasar candidate catalog of SDSS DR6 (R09), which will be helpful for the future spectroscopic observations.

We cross-matched the SDSS DR6 1 million quasar candidate catalog of R09 with the UKIDSS/LAS DR8 data by using the positional offset of 3$''$ for finding only the closest counterpart,
and obtained  97,923 sources with full detections in SDSS and UKIDSS 9 photometric bands. This SDSS-UKIDSS quasar candidate sample is much bigger than the previous one with 42,133 sources  from the UKIDSS/LAS DR3 (Peth, Ross \& Schneider 2011). Among these 97,923 sources, there are 24,878 known quasars and 73,011 unidentified quasar
candidates in SDSS DR6. 

First we checked the improvement of photometric redshift estimations using the 9-band SDSS-UKIDSS  photometric data than using the SDSS data alone. We used our photometric redshift estimation program (Wu \& Jia 2010; Wu, Zhang \& Zhou 2004) to obtain the photometric redshifts of all unidentified quasar candidates and known quasars in R90 based on the SDSS-UKIDSS data, and compared them with the photometric redshifts given in R90 and the spectral redshifts for known quasars in SDSS DR6.
In two upper panels of Fig. 2, we compare the photometric redshifts in R90 and ours for 73,011 unidentified quasar
candidates in R09, and show the histogram distribution of their differences. For 59.8\% of these unidentified
quasar candidates, the differences between two kinds of photometric redshifts are less than 0.2. However, there are still
obvious differences, especially when the photometric redshifts are smaller than 3. Comparing with our results, the photometric redshifts given in R09 are systematically larger for some low-redshift quasar candidates.
In two middle panels and two lower panels of Fig. 2, we compare the photometric redshifts given in R90
and by us for 24,878 known quasars with spectral redshifts, respectively (two middle panels are similar to
Figure 7 in Peth et al. (2011) but with more known quasars because we used the data in UKIDSS/LAS DR8). For 76.1\% of the known
quasars, R90 gave the photometric redshifts within the difference smaller than 0.2 from their spectral redshifts.  By using the SDSS-UKIDSS 9-band photometric data to estimate the photometric redshifts, such a fraction increases to 85.2\%. This significant improvement can be clearly observed from Fig. 2, and demonstrates again that by adding the near-IR photometric data to the SDSS optical data we can achieve substantially higher accuracy
in photometric redshift estimations (Wu \& Jia 2010; Wu et al. 2012). 

Next we checked the possible  star contaminations in the quasar candidate catalog of SDSS DR6 (R09), using
the Y-K/g-z quasar selection criterion. In Fig. 3 we show the distributions of  24,878 known quasars and 73,011 unidentified quasar candidates in R09 in the Y-K/g-z color-color diagram, as well as our Y-K/g-z quasar selection criterion (Wu \& Jia 2010). 
For 24,648 known $z<4$ quasars, using the  Y-K/g-z criterion can select 24,295 of them (98.6\%).
For 61,489 unidentified quasar
candidates in R09 with the photometric redshifts of $z_{ph}<2.2$ (we adopted the photometric redshifts estimated with the SDSS-UKIDSS 9-band photometric data), using the Y-K/g-z criterion can select 60,412 of them
(98.3\%). For 10,687 unidentified quasar
candidates in R09 with photometric redshift of $2.2<z_{ph}<4$, using the Y-K/g-z criterion we can select 8,934 of them
(83.6\%). Therefore, the quasar candidates selection in R09 are well consistent with our 
 Y-K/g-z selection for $z_{ph}<2.2$ quasar candidates, but there are substantial contaminations from stars for selecting $2.2<z_{ph}<4$ quasar candidates. This can be also seen from the lower panel of Fig. 3, where the green dots
below the line most probably represent the star contaminations.

To better understand the quasar selection efficiency and the star contaminations at different redshift, in Fig. 4 we plot the photometric redshift dependences of the fraction of 24,648 known $z<4$ quasars selected by
the Y-K/g-z criterion and the fraction of 72,176 unidentified quasar candidates in R09 with photometric redshifts of $z_{ph}<4$ selected by
the Y-K/g-z criterion. For known $z<4$ quasars, using the Y-K/g-z criterion can  reach the efficiency
 higher than 90\%  at almost all redshift, except for $z>3.5$. For unidentified quasar candidates, R09 selection has
the similar  efficiency (higher than 90\%)  as using the  Y-K/g-z criterion for selecting $z_{ph}<2.6$ quasars but has higher
star contaminations for selecting $z_{ph}>2.6$ quasars than using the  Y-K/g-z criterion. On may think that
the decrease of
quasar selection fraction at $z_{ph}>2.6$ (denoted by the blue dotted line in Fig. 4) is due to both the mis-identifications of quasars as stars by the  Y-K/g-z criterion and the true star contaminations. After
the deduction of the mis-identification rate of quasars as stars (which can be estimated from the known quasar selection fraction denoted as the black solid line in Fig.4) by the  Y-K/g-z criterion at different redshift,
we can obtain the possible star contamination rate in R09 at different redshift (denoted by the red dashed line in
Fig. 4). It is clear that the star contamination rate becomes substantially higher for selecting $z_{ph}>2.6$ quasars than the lower redshift ones, even up to
30\% to 40\% for selecting quasars at redshift $3<z_{ph}<3.5$. We must notice that the real star
contaminations are probably much higher than those we estimated with the  Y-K/g-z criterion. Therefore,
the star contamination rate in R09 we obtained for selecting $2.6<z_{ph}<4$ quasars could be  underestimated. 
In fact, based on the clustering study of Myers et al. (2006), the star contamination in the 'mid-z'  range of R09 was estimated to be
higher than 50\% (Richards et al. 2009b). Nevertheless, 
we believe that using the  Y-K/g-z criterion can help us to  exclude the star contaminations significantly and obtain
higher efficiency in selecting  $2.6<z_{ph}<4$  quasars.

From R09, we can obtain a list of SDSS DR6 unidentified quasar candidates with UKIDSS/LAS DR8 full detections in the YJHK bands and with the photometric redshifts (given in R09)
at $2.2\le z_{ph}(R09)\le 3.5$, which consists of  17,719
objects. However, if we adopt our photometric redshifts obtained from the 9-band SDSS-UKIDSS photometric data and use our  Y-K/g-z criterion to do further selection, such a list consists of only 7,727 quasar
candidates at $2.2\le z_{ph}\le 3.5$. The substantial decrease of the size is mainly due to the increase of photometric redshift reliability and the deduction of star contaminations by using the  Y-K/g-z criterion.
In Table 4 we list the name, photometric redshift, SDSS and UKIDSS magnitudes for these 7727 quasar candidates with our estimated photometric redshift at  $2.2\le z_{ph}\le 3.5$. We noticed that some of them have been
identified after SDSS DR6, including this work. Future spectroscopy on these unidentified quasar candidates will provide further checks to the robustness of both the quasar selection criterion and the photometric redshift estimation method.

\section{Comparisons with SDSS-III DR9 quasars}

Very recently, SDSS-III/BOSS has released the DR 9 quasar catalog, which consists of 87,822 quasars (78,086 are new and 61,931 have redshifts higher than 2.15) detected over a sky area of 3,275 deg$^2$ (Paris et al. 2012). This provides us a chance to check the effectiveness of our proposed Y-K/g-z criterion with the largest
sample of $z>2.1$ quasars currently available. 

After cross-matching the SDSS-III DR9 quasar catalog with the UKIDSS/LAS DR8 catalog, 17,999 among
87,822 quasars have available Y and K-band data, with a sampling rate of 20.5\%. 17,308 of these 17,999 quasars satisfy the Y-K/g-z selection criterion for $z<4$ quasars, with a completeness of 96.2\%. In the
upper panel of Fig. 5, we show the distributions of 17,999 SDSS-III/DR9-UKIDSS/LAS/DR8 (hereafter DR9-UKIDSS) quasars in the Y-K/g-z color-color
diagram and compare them with our proposed Y-K/g-z selection criterion for $z<4$ quasars. Similar to the upper panel of Fig. 3, this comparison also clearly
demonstrates the effectiveness in using the  Y-K/g-z selection criterion to select $z<4$ quasars. 

We also check whether the Y-K/g-z selection depends on the magnitude and redshift. In the 
middle panel of Fig. 5 we show the magnitude dependence of the selection fraction of 17,999 DR9-UKIDSS quasars by our Y-K/g-z criterion, and the normalized magnitude distributions (fraction between the number of quasars in each magnitude bin and the total number) for 17,308 quasars selected by the Y-K/g-z criterion and  for 87,822 SDSS-III/DR9 quasars.  The comparison shows that using UKIDSS data we do select optically
brighter quasars (especially those brighter than $i=20.5$) due to the limited sensitivity of UKIDSS. However,
the selection efficiency of using the Y-K/g-z criterion does not significantly depend on the magnitudes for quasars with UKIDSS data at $i<20.5$.  In the
lower panel of Fig. 5 we show the redshift dependence of the selection fraction of 17,999 DR9-UKIDSS quasars by our Y-K/g-z criterion, which clearly states that our criterion is robust for selecting $z<4$ quasars. Comparing the normalized redshift distributions (fraction between the number of quasars in each redshift bin and the total number) for 17,308 quasars selected by the Y-K/g-z criterion and  for 87,822 SDSS-III/DR9 quasars also demonstrates the similar redshift distribution of quasar sample selected by the Y-K/g-z criterion and the 
SDSS-III/DR9 quasar sample at $z<4$.

In addition,  we need to check whether using  the Y-K/g-z criterion we actually select quasars with specific colors. In Fig. 6 we show the distributions of $\Delta(g-i)$ versus redshift for 17,308 quasars selected by the Y-K/g-z criterion and  for 87,822 SDSS-III/DR9 quasars. Obviously they significantly overlap at $z<4$. From the distribution of median  $\Delta(g-i)$ values  in each redshift bin  the quasars selected by the Y-K/g-z criterion have slightly
redder $g-i$ color than the SDSS-III/DR9 quasars at $z<4$. The median $\Delta(g-i)$ value for 17,308 quasars selected by the Y-K/g-z criterion is 0.088, which is slightly larger that the median value
0.035 for  87,822 SDSS-III/DR9 quasars. This is mainly because using the UKIDSS near-IR data we
mostly select quasars with $i<20.5$ (see the middle panel of Fig. 5). Most SDSS-III/DR9  quasar with
$i>20.5$ have smaller or negative $\Delta(g-i)$ values. We also noticed that the SDSS-III/DR9 quasars with
$i<20.5$ have median $\Delta(g-i)$ value of 0.080, which is close to median value 0.088 of the Y-K/g-z selected quasars.

Finally we check whether using  the Y-K/g-z criterion can help  us to select more BAL quasars. In
SDSS-III/DR9 quasar catalog, 7,533 quasars are found to be BAL quasars after the visual inspection. 2,173 of them
have UKIDSS/LAS DR8 data and 1,974 quasars satisfy the Y-K/g-z criterion (With a percentage of 90.8\%). 
This fraction is only slightly smaller than 96.2\% for the Y-K/g-z selection of all DR9-UKIDSS quasars, 
implying that using the  Y-K/g-z criterion we can also efficiently select BAL quasars. We also
noticed that 3 quasars in our 36 newly-discovered quasars are BAL quasars (see section 2), which is consistent with the
fraction of 7,533 BAL quasars in 87,822 SDSS-III/DR9 quasars, though with small number statistics.

\section{Selecting red quasars and type II quasars with the Y-K/g-z criterion}
In this section we check whether  we can also efficiently select red quasars and type II quasars by using the Y-K/g-z criterion. 

Glikman et al. (2007,2012)  presented a sample of 128 red quasars, with redshifts up to 3.05, selected from the FIRST-2MASS radio and near-IR surveys. These quasars with red color, J-K$>$1.7, are thought to be transient objects between heavily obscured quasars and normal blue quasars. Urrutia et al. (2009) also identified 57 red quasars with J-K$>$1.3 from 122 candidates selected from FIRST, 2MASS and SDSS, with a high fraction of BAL quasars. By cross-matching with the UKIDSS LAS/DR8 catalog, we get 26 and 16 red quasars with Y and K-band detections from these two samples, respectively. We plot them in the Y-K/g-z color-color diagram (see the upper panel of Fig. 7) and find that 23 among 26 red quasars, and 14 among 16 red quasars in these two samples satisfy the Y-K/g-z criterion. The selection efficiency is 88.5\% and 87.5\% for these two samples, respectively. This comparison clearly demonstrates the high efficiency of selecting red quasars with the Y-K/g-z criterion.

Because the SDSS-III/DR9 quasar catalog does not provide the information about how many type II quasars are included in it,  we use the existed type II quasar catalog in SDSS to check the efficiency of selecting type II quasars with the Y-K/g-z criterion. Using the catalog provided by
Reyes et al. (2010), which includes 887 optically selected  $0.3<z<0.83$ type II SDSS quasars, we find that 282 among 887 type II  quasars have available Y and K-band data from UKIDSS LAS/DR8. We plot them in the Y-K/g-z color-color diagram, in comparison with the  Y-K/g-z selection criterion (see the lower panel of Fig. 7). 272 of these 282 type II quasars (with a percentage of 96.5\%) satisfy our  Y-K/g-z criterion. This selection fraction is similar as 96.2\% for the Y-K/g-z selection of all DR9-UKIDSS quasars, which indicates that using  the Y-K/g-z criterion we can also efficiently discover type II quasars. 

In Fig. 7, we also plot the predicted color tracks for type I quasars and type II quasars at different redshifts (up to 4.3, to the right side in Fig. 7), using the related spectral templates from Polletta et al. (2007). We can clearly see that although type II quasars have redder Y-K and g-z colors than type I quasars, using the Y-K/g-z criterion we can efficiently select both type I and type II quasars with redshifts up to 4.

\section{Can we use the Y-K/g-z criterion to select more $2.2<z<3.5$ quasars?}

Although our above test with the recently released SDSS-III/DR9 quasar catalog does indicate the effectiveness of using the Y-K/g-z critirion in selecting $z<4$ quasars, one question still needs to be addressed. Can we use the Y-K/g-z criterion to select additional $2.2<z<3.5$ quasars which SDSS-III/BOSS does not select? 

To answer this question, we should find a sky area which covered both by SDSS-III/BOSS and UKIDSS, and compare the quasar candidates selected by the SDSS-III/BOSS and Y-K/g-z selection criterion. Thanks to Adam Myers in the SDSS-III/BOSS team, who kindly provides us the photometric catalog of a 15 square-degree region in SDSS Stripe 82 (with $36^o<RA<42^o$ and $-1.25^o<Dec<1.25^o$), which has relatively complete spectroscopic observations on SDSS-III/BOSS quasar targets. By cross-matching the 90,922 SDSS sources in this area with UKIDSS and selecting sources with SDSS photometric parameters type=6, r$<$21.85 and g$<$22 (the same as in SDSS-III/BOSS, r and g magnitudes are Galactic extinction corrected), we get 24,627 point sources with UKIDSS/LAS Y and K-band detections. Among them, 21,715 were unidentified while 2,912 were identified (including 743 quasars and 2,169 stars). 135 of 743 identified quasars are in the redshift range between 2.2 and 3.5. 130 of them were included in the SDSS-III/DR9 quasar catalog and 5 of them were identified by other BOSS ancillary programs. We also found that 712 of 743 identified quasars, including 126 of 135 $2.2<z<3.5$ quasars, satisfy the Y-K/g-z selection criterion.

Among 21,715 unidentified sources, 340 of them satisfy the Y-K/g-z selection criterion. The photometric redshifts of these 340 quasar candidates were estimated based on their SDSS and UKIDSS data, and 140 sources are found to have photometric redshifts between 2.2 and 3.5. If we further require their $\chi^2$ values of photometric redshift estimations (Wu \& Jia 2010; Wu et al. 2004; Weinstein et al. 2004) smaller than 10, which is satisfied by 94\% of 743 identified quasars in this sky area, the number of $2.2<z<3.5$ quasar candidates selected by the Y-K/g-z criterion becomes 86. Even if we require the $\chi^2$ values smaller than 6, which is satisfied by 80\% of 743 known quasars in this area, the number of Y-K/g-z selected $2.2<z<3.5$ quasar candidates becomes 52. Because there are 470 known $2.2<z<3.5$ quasars (421 are  SDSS-III/DR9 quasars) in this sky area and 135 of them (mostly brighter ones) have UKIDSS Y and K-band detections, the Y-K/g-z selected additional $2.2<z<3.5$ quasar candidates may add at least 10\% to the total number of SDSS-III/BOSS $2.2<z<3.5$ quasars. However, whether these candidates are real $2.2<z<3.5$ quasars still needs to be confirmed by the future spectroscopic observations. Therefore, with this check we believe that SDSS-III/BOSS has selected most $2.2<z<3.5$ quasars, and the Y-K/g-z selection may add about 10\% additional $2.2<z<3.5$ quasars in the UKIDSS surveyed area.

\section{Discussion}
We have presented the spectroscopic observations on 43 bright quasar candidates selected from R09, which have photometric redshifts at $2.2<z_{ph}<3.5$ estimated from the 9-band SDSS and UKIDSS photometric data  and satisfy our Y-K/g-z criterion, and successfully identified 36 of them to be
real quasars with redshifts between 2.1 and 3.4. 
The high efficiency of spectroscopic identifications 
provides further support for discovering more quasars at intermediate redshifts based on the optical and near-IR color selections. We also found substantial improvement of photometric redshift estimation
from using the 9-band SDSS-UKIDSS data than using the SDSS data alone. We investigated the star contamination rate of quasar candidates in R09, which could be much higher  for selecting quasars at photometric redshift of 
$3<z_{ph}<3.5$ than the lower redshift ones ($z<2.2$). By using our photometric redshifts estimated from the SDSS and UKIDSS photometric data and the  Y-K/g-z criterion to exclude the star contaminations, we obtained a catalog of 7727 SDSS-UKIDSS unidentified
quasar candidates with  photometric redshifts at
$2.2<z_{ph}<3.5$. The ongoing and future spectroscopic observations, such as SDSS-III/BOSS(Eisenstein et al. 2011), will
provide further check to the robustness of this catalog, though the UKIDSS near-IR data were not used for
selecting the majority of quasar candidates in BOSS (Ross et al. 2012). 

Using the recently released
SDSS-III/DR9 quasar catalog and UKIDSS/LAS DR8 data, we find that 96.2\% of UKIDSS detected DR9 quasars, including 90.8\% of BAL quasars, satisfy the  Y-K/g-z criterion. This provides further support to this criterion for selecting $z<4$ quasars, including BAL quasars. We also check the efficiency of using the  Y-K/g-z criterion to select red quasars and type II quasars with some available samples, and find that about 88\% of
red quasars with $z<3.05$ and 96.5\% of type II quasars with $z<0.83$ satisfy the Y-K/g-z criterion. These results, together with the predicted color tracks by using different spectral templates of quasars,  support the robustness of using the Y-K/g-z criterion to discover both unobscured and obscured quasars. Our test in a small sky area of SDSS Stripe 82 also proves that with the Y-K/g-z selection criterion we may add about 10\% additional $2.2<z<3.5$ quasars to the SDSS-III/BOSS quasars in the UKIDSS surveyed area.

Since UKIDSS only covers a very limited sky area, we still need much deeper optical/near-IR photometry in a larger sky area for taking the full advantages of the optical/near-IR color for selecting quasars,  
especially for  $z>2.2$ quasars.  
The recently released Wide-field Infrared Survey Explorer (WISE) all-sky data (Wright et al. 2010) also
provided abundant photometric data in the near(middle)-IR bands, which will be very helpful for quasar selections (Wu et al. 2012; Stern et al. 2012; Edelson \& Malkan 2012; Yan et al. 2013)
Fortunately, several ongoing  optical and near-IR photometric sky surveys will also provide
us further oppotunities to apply our  optical/near-IR color selections of quasars to larger
and deeper fields. In addition to SDSS III (Eisenstein et al. 2011), which has taken 2,500 deg$^2$ 
further imaging
in the south galactic cap, the SkyMapper (Keller et al. 2007) and Dark Energy Survey (DES; The Dark Energy Survey Collaboration 2005) will also present the multi-band optical 
photometry in 20,000/5,000 deg$^2$ of the southern sky, with the magnitude limit of 22/24 mag in $i$-band,
respectively. The Visible and Infrared Survey Telescope for Astronomy (VISTA; Arnaboldi et al. 2007) is
carrying out the
VISTA Hemisphere Survey (VHS) in the near-IR YJHK bands for 20,000 deg$^2$ of the southern sky with
a magnitude limit at K=20.0, which is about five and two magnitude deeper than the Two Micron ALL
Sky Survey (2MASS; Skrutskie et al. 2006)
and UKIDSS/LAS limits (Lawrence et al. 2007), respectively. Therefore, the optical and near-IR photometric data
obtained with these ongoing surveys will provide us a large database for quasar selections. Needless to say,
 the ongoing Panoramic Survey Telescope \& Rapid Response System (Pan-STARRS;
Kaiser et al. 2002) and the future Large Synoptic Survey Telescope (LSST; Ivezic et al. 2008) will also
provide us with multi-epoch photometry in multi-bands covering a large area of the sky, which will undoubtedly  help us to construct a much larger  sample of 
quasars based on both optical/near-IR colors and variability features.

On the other hand,  the spectroscopic 
observations are still crucial to determine the quasar nature and redshifts for the quasar candidates selected from the optical/near-IR colors.
The ongoing SDSS-III/BOSS is expected to obtain
the spectra of 150,000 quasars at  $2.2<z<4$ (Eisenstein et al. 2011; Ross et al. 2011). We believe that many      
  $2.2<z<3.0$ quasars, including the candidates we listed in this paper,  should be 
spectroscopically identified by BOSS. In addition, 
the Chinese GuoShouJing telescope (LAMOST; Su et al. 1998; Cui et al. 2012; Zhao et al. 2012), a spectroscopic telescope with 4000 fibers, which is
currently in the final stage of commissioning and will start the regular spectroscopic survey in the fall of 2012, 
is also aiming at discovering 0.3 million quasars from 1 million candidates
with magnitudes bright than $i=20.5$ in the next 5 years (Wu et al. 2010a,b; Wu 2011). 
By using the optical/near-IR colors, we hope the larger input catalogs of reliable quasar candidates
will be provided to these quasar surveys for future spectroscopic observations. We expect that a much larger and more complete quasar sample covering a wider range of
 redshift  will be constructed in the near future.  In addition, the results we presented in this work may
be also helpful to the future spectroscopic surveys of quasars, like those in eBOSS and Big BOSS (Schlegel et al. 2011).

\acknowledgments 
We thank the referee, Nicholas Ross, for very constructive suggestions to improve the paper, and Adam Myers for providing us a photometric
catalog of an eBOSS test region in SDSS Stripe 82.
This work was supported by the National Natural Science Foundation of China (grant No. 11033001) and
 by the Open Project Program of the Key Laboratory of Optical Astronomy, NAOC, CAS.

Funding for the SDSS and SDSS-II has been provided by the Alfred P. Sloan Foundation, the Participating Institutions, the National Science Foundation, the US Department of Energy, the National Aeronautics and Space Administration, the Japanese Monbukagakusho, the Max Planck Society and the Higher Education Funding Council for England. The SDSS web site is http://www.sdss.org/.

The SDSS is managed by the Astrophysical Research Consortium for the Participating Institutions. The Participating Institutions are the American Museum of Natural History, Astrophysical Institute Potsdam, University of Basel, University of Cambridge, Case Western Reserve University, University of Chicago, Drexel University, Fermilab, the Institute for Advanced Study, the Japan Participation Group, Johns Hopkins University, the Joint Institute for Nuclear Astrophysics, the Kavli Institute for Particle Astrophysics and Cosmology, the Korean Scientist Group, the Chinese Academy of Sciences (LAMOST), Los Alamos National Laboratory, the Max-Planck-Institute for Astronomy (MPIA), the Max-Planck-Institute for Astrophysics (MPA), New Mexico State University, Ohio State University, University of Pittsburgh, University of Portsmouth, Princeton University, the United States Naval Observatory and the University of Washington.

{\it Facilities:} \facility{Sloan (SDSS)}, \facility{UKIDSS},\facility{2.16m/NAOC}

\begin{table}
{\scriptsize 
\caption{Procedures of target selection}
\begin{tabular}{ccl}
\hline \noalign{\smallskip}
Step & Number of candidates & Description\\
\hline \noalign{\smallskip}
1 & 1,015,082 & All quasar candidates in R09\\
2 & 925,899 & Excluding known quasars in SDSS DR6\\
3 & 8,845 & Brighter than i=18.5\\
4 & 2,149 &  Photometric redshift $z_{phot}>2.2$ and probability $z_{prob}>0.5$ in R09\\
5& 401 & With UKIDSS LAS DR7 data\\
6& 126 & Satisfying Y-K$>$0.46(g-z)+0.82 selection criterion\\
7& 108 & Excluding 17 quasars and 1 white dwarf identified after SDSS DR6\\
8& 93 & Excluding 15 candidates with SDSS-UKIDSS based photometric redshifts smaller than 2\\
9& 105 & Adding 12 candidates not included in UKIDSS/LAS DR 7 but included in UKIDSS/LAS DR8\\
\noalign{\smallskip} \hline \noalign{\smallskip}
\end{tabular}\\}
\end{table}

\begin{table}
{\scriptsize \caption{Parameters and observation details of 43 quasar candidates}
\setlength{\tabcolsep}{0.9pt}
\begin{tabular}{cccccccccccccccc}
\hline \noalign{\smallskip}
Name & Date & Exposure & u &g& r& i& z& Y& J& H& K&  $z_{ph}(R09)$ &$ z_{ph}$&Result& $ z_{sp}$\\
(SDSS J) & & (s)  &    &   &  &  & & & & & & & & &  \\
\noalign{\smallskip} \hline \noalign{\smallskip}
075746.08+232054.2&2012-03-13&3600&	19.06&18.46&18.48&18.47&18.27&17.69&17.39&16.82&16.02&2.365&	2.375&	quasar&2.532\\
081545.72+264847.1&2012-04-14&	1800&18.27&17.14&17.06&17.04&16.87&16.24&15.86&15.5&15.29&	2.755&2.825&	F star	&	\\
081617.55+225604.5&2012-05-15&2400&	19.95&18.63&18.43&18.35&18.35&17.74&17.37&16.92&16.57&	2.905&2.875&	quasar&2.931\\
083255.70+004710.1&2012-03-13&3600&	19.80&18.40&18.29&18.30&18.28&17.89&17.36&16.92&16.54&2.905&	2.875&	quasar&2.919\\
084659.42+253940.9&2012-04-16&	3600&19.75&18.44&18.46&18.43&18.25&17.46&17.03&16.53&15.99&2.795&2.875&	quasar&2.892\\
085152.98+091808.5&2012-05-16&3600&	18.65&18.13&17.95&17.94&17.79&17.15&16.95&16.54&15.85&	2.505&2.575&	low S/N&\\
085825.51+283258.5&2012-03-13&3600&20.54&18.73&18.44&18.50&18.49&17.68&17.20&16.74&16.15&	3.535&	2.975&	quasar&3.226\\
090233.19+034131.2&2012-05-17&3600&18.49&17.92&17.79&17.77&17.6&17.06&16.88&16.39&15.63&2.465&	2.575&	quasar&2.532\\
090827.71+011322.5&2012-04-15&	2400&18.33&17.64&17.54&17.47&17.29&16.83&16.51&16.07&15.26&	2.565&2.325&	quasar&2.083\\
091756.58+100836.7&2012-04-16&	2400&18.54&17.72&17.68&17.67&17.58&16.9&16.55&16.1&15.67&2.695&2.675&	quasar&2.760\\
091857.36+025205.4&2012-04-15&	3000&19.39&17.9&17.79&17.72&17.61&17.06&16.75&16.39&15.9&	2.905&2.875&	quasar&2.789\\
092021.02-020113.7&2012-03-14&3600&19.96&18.58&18.51&18.46&18.44&17.76&17.45&17.15&16.71&	2.905&	2.875&	quasar&2.826\\
093655.11+305855.3&2012-05-17&2700&20.37&18.55&18.45&18.49&18.37&17.50&17.31&16.8&16.37&2.945	&2.875&	quasar&3.005\\
094137.09+102650.9&2012-04-16&	2700&18.67&18.04&17.90&17.9&17.75&17.01&16.77&16.38&15.84&	2.535&2.625&	quasar&2.584\\
095118.43+083959.9&2012-03-13&3600&	18.90&18.25&18.21&18.19&18.02&17.50&17.18&16.64&15.85&	2.365&	2.375&	quasar&2.523\\
100834.73-022302.5&2012-04-14&3600&19.93&18.13&17.91&17.88&17.85&17.39&16.95&16.54&16.01&3.005&2.975&	quasar&3.086\\
103301.49+065106.5&2012-05-16&2700&18.73&18.03&17.91&17.84&17.69&17.15&16.88&16.45&16.05&2.565	&2.625&	quasar&2.637\\
114608.05+094216.5&2012-05-15&5400&19.54&18.49&18.49&18.38&17.96&17.05&16.83&16.22&15.35&	2.615&	2.475&	quasar&2.580\\
115531.45-014611.9&2012-03-13&6000	&23.25&19.36&18.91&18.50&18.50&17.94&17.39&17.00&16.36&	3.495&	3.350&	quasar(BAL)&3.196\\
121510.62+142834.5&2012-04-15&	3600&19.20&18.45&18.34&18.33&18.15&17.47&17.20&16.75&16.11&	2.595&2.675&	quasar&2.480\\
122043.86+011122.1&2012-05-17&3600&	19.21&18.39&18.38&18.27&18.00&17.20&16.86&16.31&15.61&	2.395&2.725&	quasar&2.565\\
122619.73+104953.5&2012-04-15&3600&19.06&18.35&18.27&18.27&18.10&17.49&17.23&16.78&16.12&2.565&2.625&		quasar&2.375\\
124605.36+071128.2&2012-03-14&3600&19.20&18.56&17.82&17.52&17.33&16.88&16.55&15.97&15.25&3.435&	3.550&quasar&2.044\\
125934.29+075200.7&2012-05-16&2700&	18.29&17.78&17.67&17.68&17.63&17.2&16.93&16.45&15.93&	2.475&2.625&	quasar&2.370\\
130318.32+030809.4&2012-04-14&2700&18.39&17.65&17.58&17.47&17.35&16.6&16.25&15.87&15.37&2.595&2.675&	quasar&2.664	\\
131008.67+084405.0&2012-04-16&	2700&18.48&17.85&17.73&17.72&17.63&17.15&16.92&16.43&15.85&	2.535&2.625&	quasar&2.232\\
135942.50+022426.0&2012-03-14&3600&24.10&18.82&18.33&18.29&18.18&17.69&17.26&16.85&16.33&	3.475&	3.350&	quasar(BAL)&3.265\\
142405.57+044105.5&2012-05-16&3600&	18.72&18.17&18.01&18.01&17.83&17.36&17.09&16.50&15.89&	2.465&2.625&	quasar(BAL)&2.232\\
142543.33+024759.8&2012-04-15&	2700&18.48&17.79&17.77&17.75&17.67&16.94&16.65&16.16&15.70&	2.605&2.675&	quasar&2.689\\
142854.09+132259.0&2012-05-17&5400&	20.89&19.46&18.94&18.37&18.10&17.26&16.86&16.38&15.83&	2.735&2.925&	quasar&3.093\\
144526.15+023906.8&2012-05-15&3600&	19.05&18.00&18.00&17.95&17.78&17.24&16.91&16.46&15.98	&2.695&	2.675&	quasar&2.706\\
145230.38+130227.3&2012-05-17&2700&	18.26&17.77&17.65&17.68&17.51&16.81&16.58&16.03&15.33&	2.465&2.525&	quasar&2.468\\
151321.18+012502.2&2012-04-16&	3600&19.45&18.18&18.28&18.17&18.05&17.34&17.01&16.36&15.81&	2.865&2.825&	quasar&2.753\\
152808.87+005211.8&2012-04-16&	3600&18.97&18.11&18.09&18.07&17.95&17.20&16.89&16.54&16.12&	2.675&2.675&	quasar&2.610\\
153303.54+064032.9&2012-04-15&	3600&22.74&18.99&18.43&18.35&18.26&17.84&17.26&16.81&16.15&	3.405&3.350&		quasar&3.422\\
153319.44+043257.3&2012-04-14&5400&18.81&18.10&17.99&17.96&17.74&17.28&17.08&16.74&15.95&	2.535&2.575&	low S/N&\\
153515.55+291038.5&2012-04-14&2700&19.08&17.71&17.50&17.36&17.30&16.79&17.29&16.17&16.10&2.905&2.775&	F star&\\
153550.13+063352.8&2012-05-17&3600&	19.08&18.43&18.32&18.32&18.12&17.90&17.57&17.15&16.37&	2.535&2.275&	low S/N&\\
153551.88+044416.4&2012-05-16&3600&	18.73&18.19&18.03&18.01&17.92&17.79&17.50&17.02&16.13&	2.535&2.275&	quasar&2.377\\
153951.05+020133.8&2012-05-16&3600&	18.90&18.31&18.24&18.21&18.08&17.32&17.11&16.78&16.13&	2.365&2.575&	quasar&2.569\\
154503.23+015614.7&2012-05-16&2700&	18.36&17.90&17.64&17.65&17.51&17.03&16.71&16.21&15.42&	2.505&2.225&	low S/N&\\
162352.69+230119.6&2012-05-15&5400&	20.15&19.21&18.79&18.45&17.96&17.20&16.57&16.08&15.76&	2.715&	2.925&	low S/N&\\
162620.89+282924.7&2012-04-15&	3600&19.20&18.42&18.32&18.33&18.17&17.39&17.04&16.81&16.30&	2.605&2.675&	quasar&2.534\\
\noalign{\smallskip} \hline \noalign{\smallskip}
\end{tabular}\\
Note: The SDSS ugriz magnitudes are given in AB system 
and the UKIDSS YJHK magnitudes are given in Vega system. $ z_{ph1}$, $ z_{ph2}$ and $ z_{sp}$
are the photometric redshifts obtained from the 5-band SDSS data by Richards et al. (2009) and obtained from the 9-band SDSS-UKIDSS
data by us, and the spectral redshifts from our observations, respectively. }
\end{table}

\begin{table}
\caption[]{Spectral parameters and black hole masses of 36 new quasars\label{mbh}}
\setlength{\tabcolsep}{0.9pt}
\small
 \begin{tabular}{ccccccccccccccc}
  \hline\noalign{\smallskip}
Name &  redshift & slope$^a$ & 
$\log (L_{1350})$ & FWHM(CIV) & $\log (M_{BH})$ &
$\log (L_{\rm bol})$ & $\log (R_{\rm EDD})$  \\
 (SDSS J)& & & (erg/s) & (km/s) & 
($M_\odot$) &(erg/s) & \\
  \hline\noalign{\smallskip}
J075746.08+232054.2  &   2.532$\pm$0.007 &    -1.59   &   46.39 &    6000       &    9.48 &    47.06  &   -0.53 \\
081617.55+225604.5   &   2.931$\pm$0.007 &    -0.23   &   46.45 &   11538$^b$   &   10.08 &    47.11  &   -1.07 \\
083255.70+004710.1   &   2.919$\pm$0.007 &    -0.95   &   46.43 &    4778       &    9.31 &    47.10  &   -0.31 \\
084659.42+253940.9   &   2.892$\pm$0.029 &    -1.80   &   46.54 &    4073       &    9.22 &    47.20  &   -0.12 \\
085825.51+283258.5   &   3.226$\pm$0.008 &    -1.32   &   46.74 &    6038       &    9.68 &    47.41  &   -0.37 \\
090233.19+034131.2   &   2.532$\pm$0.028 &    -0.39   &   46.30 &    7817$^b$   &    9.66 &    46.96  &   -0.80 \\
090827.71+011322.5   &   2.083$\pm$0.081 &    -1.58   &   46.82 &    8314       &    9.99 &    47.48  &   -0.61 \\
091756.58+100836.7   &   2.760$\pm$0.026 &    -2.44   &   46.86 &    6797       &    9.84 &    47.52  &   -0.42 \\
091857.36+025205.4   &   2.789$\pm$0.038 &    -0.71   &   46.52 &    6306       &    9.59 &    47.18  &   -0.51 \\
092021.02-020113.7   &   2.826$\pm$0.025 &    -3.27   &   46.61 &    6278       &    9.64 &    47.27  &   -0.47 \\
093655.11+305855.3   &   3.005$\pm$0.012 &     0.01   &   46.44 &    4699       &    9.30 &    47.10  &   -0.29 \\
094137.09+102650.9   &   2.584$\pm$0.024 &    -0.99   &   46.54 &    4779$^b$   &    9.36 &    47.20  &   -0.26 \\
095118.43+083959.9   &   2.523$\pm$0.006 &    -0.82   &   46.27 &    6448       &    9.48 &    46.94  &   -0.65 \\
100834.73-022302.5   &   3.086$\pm$0.012 &    -2.04   &   46.39 &    5209       &    9.36 &    47.05  &   -0.41 \\
103301.49+065106.5   &   2.637$\pm$0.027 &    -1.60   &   46.46 &    6817$^b$   &    9.63 &    47.13  &   -0.61 \\
114608.05+094216.5   &   2.580$\pm$0.020 &    -0.90   &   46.11 &    6631       &    9.42 &    46.77  &   -0.75 \\
115531.45-014611.9   &   3.196$\pm$0.047 &    -1.87   &   46.29 &    8483$^b$   &    9.73 &    46.96  &   -0.88 \\
121510.62+142834.5   &   2.480$\pm$0.073 &    -2.20   &   46.62 &    7208       &    9.76 &    47.29  &   -0.58 \\
122043.86+011122.1   &   2.565$\pm$0.042 &     1.00   &   45.95 &    3379$^b$   &    8.75 &    46.62  &   -0.24 \\
122619.73+104953.5   &   2.375$\pm$0.046 &    -2.06   &   46.54 &    7050       &    9.70 &    47.20  &   -0.60 \\
124605.36+071128.2   &   2.044$\pm$0.019 &    -0.10   &   46.53 &   10662$^b$   &   10.06 &    47.19  &   -0.96 \\
125934.29+075200.7   &   2.370$\pm$0.030 &    -2.02   &   46.91 &   11521$^b$   &   10.33 &    47.57  &   -0.85 \\
130318.32+030809.4   &   2.664$\pm$0.037 &    -1.91   &   46.84 &    5772       &    9.69 &    47.50  &   -0.29 \\
131008.67+084405.0   &   2.232$\pm$0.050 &    -1.57   &   46.55 &    7262       &    9.73 &    47.21  &   -0.62 \\
135942.50+022426.0   &   3.265$\pm$0.014 &    -0.42   &   46.77 &    8215$^b$   &    9.96 &    47.44  &   -0.62 \\
142405.57+044105.5   &   2.232$\pm$0.050 &    -1.05   &   46.56 &   11934$^b$   &   10.17 &    47.22  &   -1.05 \\
142543.33+024759.8   &   2.689$\pm$0.035 &    -2.51   &   46.88 &    5984       &    9.74 &    47.55  &   -0.30 \\
142854.09+132259.0   &   3.093$\pm$0.015 &     0.02   &   46.10 &    5404$^b$   &    9.24 &    46.77  &   -0.57 \\
144526.15+023906.8   &   2.706$\pm$0.017 &    -1.31   &   46.37 &    6040       &    9.48 &    47.03  &   -0.55 \\
145230.38+130227.3   &   2.468$\pm$0.015 &    -0.79   &   46.31 &    7693       &    9.66 &    46.98  &   -0.78 \\
151321.18+012502.2   &   2.753$\pm$0.035 &    -1.31   &   46.10 &    7759$^b$   &    9.55 &    46.77  &   -0.89 \\
152808.87+005211.8   &   2.610$\pm$0.014 &    -0.77   &   46.17 &   12187$^b$   &    9.98 &    46.83  &   -1.25 \\
153303.54+064032.9   &   3.422$\pm$0.021 &    -3.41   &   46.89 &   12183       &   10.36 &    47.56  &   -0.91 \\
153551.88+044416.4   &   2.377$\pm$0.025 &    -1.43   &   46.35 &    9027$^b$   &    9.82 &    47.01  &   -0.90 \\
153951.05+020133.8   &   2.569$\pm$0.028 &    -1.88   &   46.29 &    8925       &    9.78 &    46.96  &   -0.92 \\
162620.89+282924.7   &   2.534$\pm$0.034 &    -0.85   &   46.48 &    7456       &    9.72 &    47.15  &   -0.67 \\
  \noalign{\smallskip}\hline
\end{tabular}   \\
Note:    $^a$ The slope of the fitted power-law continuum. 
         $^b$ Only 1 Gaussian is fitted to the whole CIV line profile.
\end{table}

\begin{table}
 \caption{A catalog of 7727 SDSS-UKIDSS quasar candidates with $2.2\le z_{ph} \le 3.5$ selected from R09}
\setlength{\tabcolsep}{1.9pt}
\begin{tabular}{cccccccccccc}
\hline \noalign{\smallskip}
Name & $z_{ph}(R09)$ & $z_{ph}$ & u &g& r& i& z& Y& J& H& K\\
(SDSS J) & &  &    &   &  &  & & & & &   \\
\noalign{\smallskip} \hline \noalign{\smallskip}
 000005.95+145310.1   &    2.255    &    2.725   &     21.31   &     20.64    &    20.45    &    20.19   &     19.93       & 19.41      &  18.83     &   18.51    &    17.76\\
 000035.59-003146.1    &    2.255    &    2.725    &    21.63    &    21.04     &   20.83     &    20.4     &   20.36       & 19.19     &   18.92    &    18.31    &    17.51\\
 000041.87-001207.3  &      2.905  &      2.925    &    21.02    &    19.62   &     19.44    &    19.32   &     19.19      & 18.45 &       18.14   &     17.56   &     16.86\\
 000050.59+010959.1    &    2.605    &    2.575      &  19.85      &  19.08    &    19.02   &     19.09     &   18.89       & 18.33     &   18.17   &     17.69    &     16.9\\
 000201.15+001707.4   &     2.465   &     2.475    &    21.48   &     20.77     &   20.69    &    20.65    &    20.18       & 19.71      &  19.36    &    18.77      &  17.68\\
\noalign{\smallskip} \hline \noalign{\smallskip}
\end{tabular}\\
Note: The SDSS ugriz magnitudes are given in AB system 
and the UKIDSS YJHK magnitudes are given in Vega system. $ z_{ph}(R09)$ and $ z_{ph}$ 
are the photometric redshifts obtained from the 5-band SDSS data by Richards et al. (2009) and obtained from the 9-band SDSS-UKIDSS
data by us, respectively. Only a portion of the table is shown here. The whole table is available in the electronic version. 
\end{table}

\newpage
\begin{figure}
\plotone{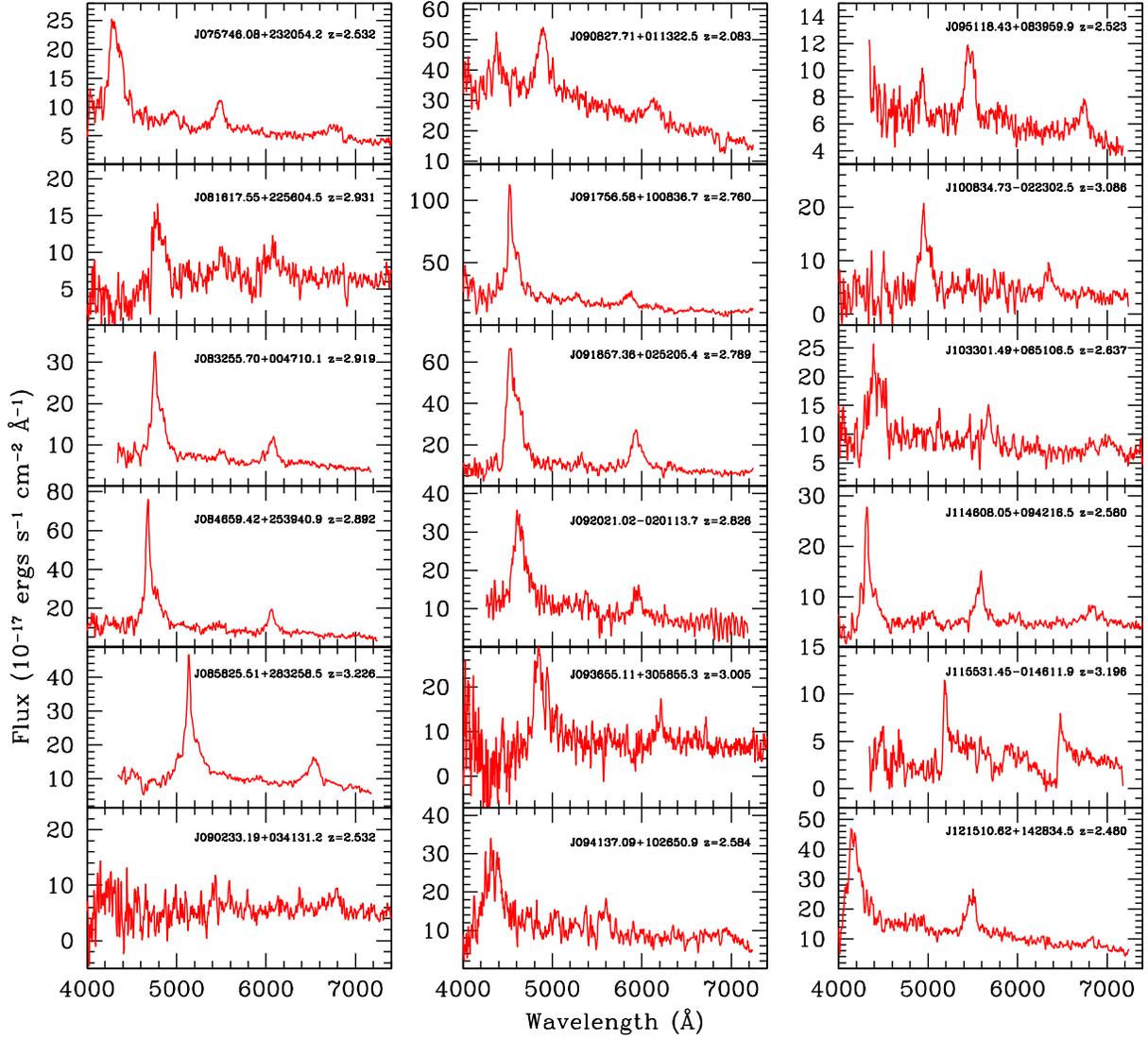}
\caption{ The spectra of the 36 new quasars at $2.2<z<3.5$ identified with the BFOSC of the Xinglong 2.16m telescope, NAOC.
 The strongest emission line
in each spectrum is Ly$\alpha$+$\rm N\,{\small V}$.}
\end{figure}
\begin{figure}
\plotone{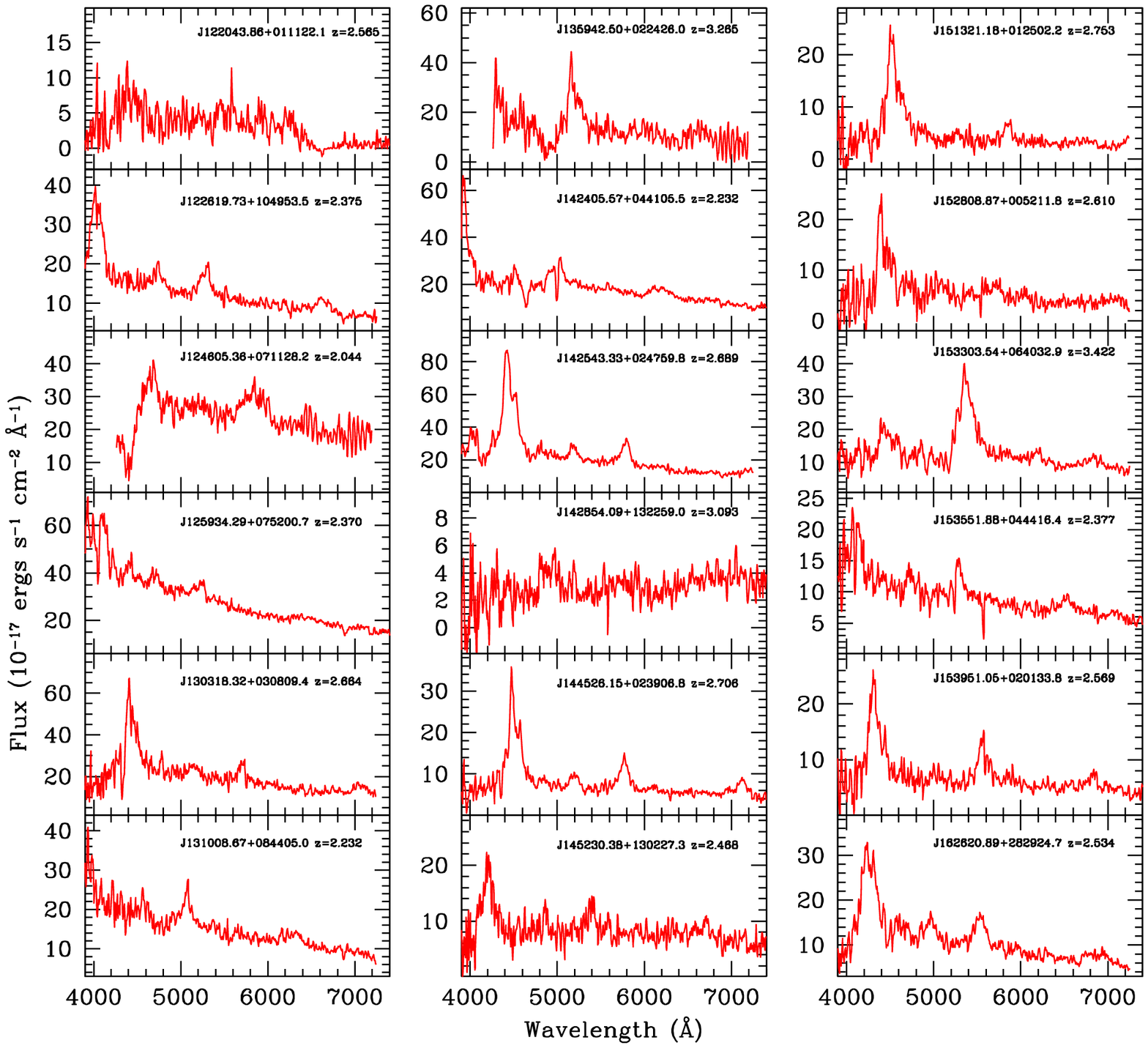}
Fig. 1: (Continued)
\newpage
\end{figure}

\begin{figure}
\plotone{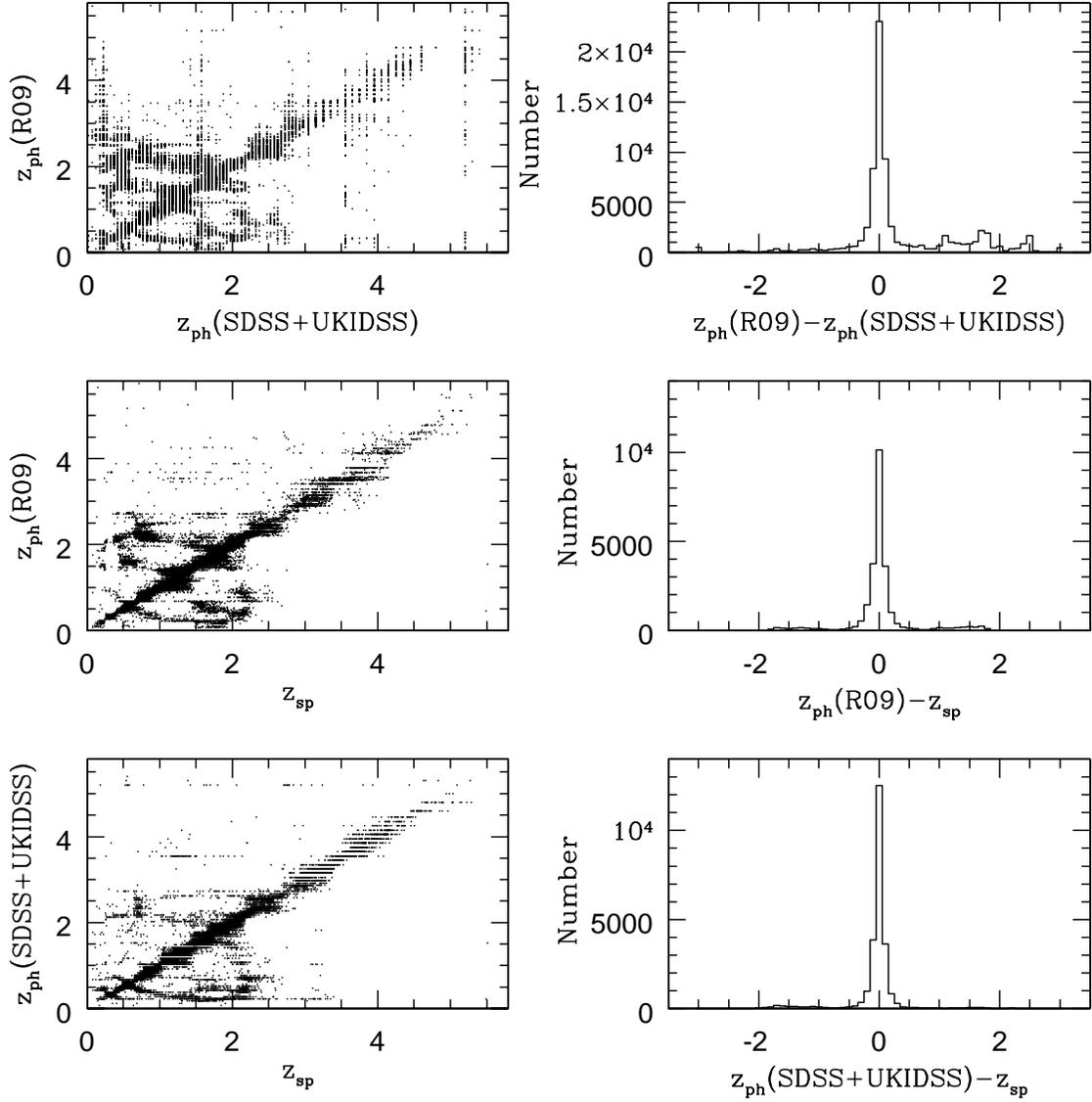}
\caption{Upper panels: Comparison of the photometric redshifts in R90 and ours based on SDSS-UKIDSS 9-band data for 73,011 unidentified quasar
candidates in R09 and  the histogram distribution of their differences. Middle panels: Comparison of the photometric redshifts in R90 with the spectral redshifts for 24,878 known quasars
and  the histogram distribution of their differences. Lower panels: Comparison of the photometric redshifts given by us with the spectral redshifts for 24878 known quasars
and  the histogram distribution of their differences.} 
\end{figure}

\begin{figure}
\plotone{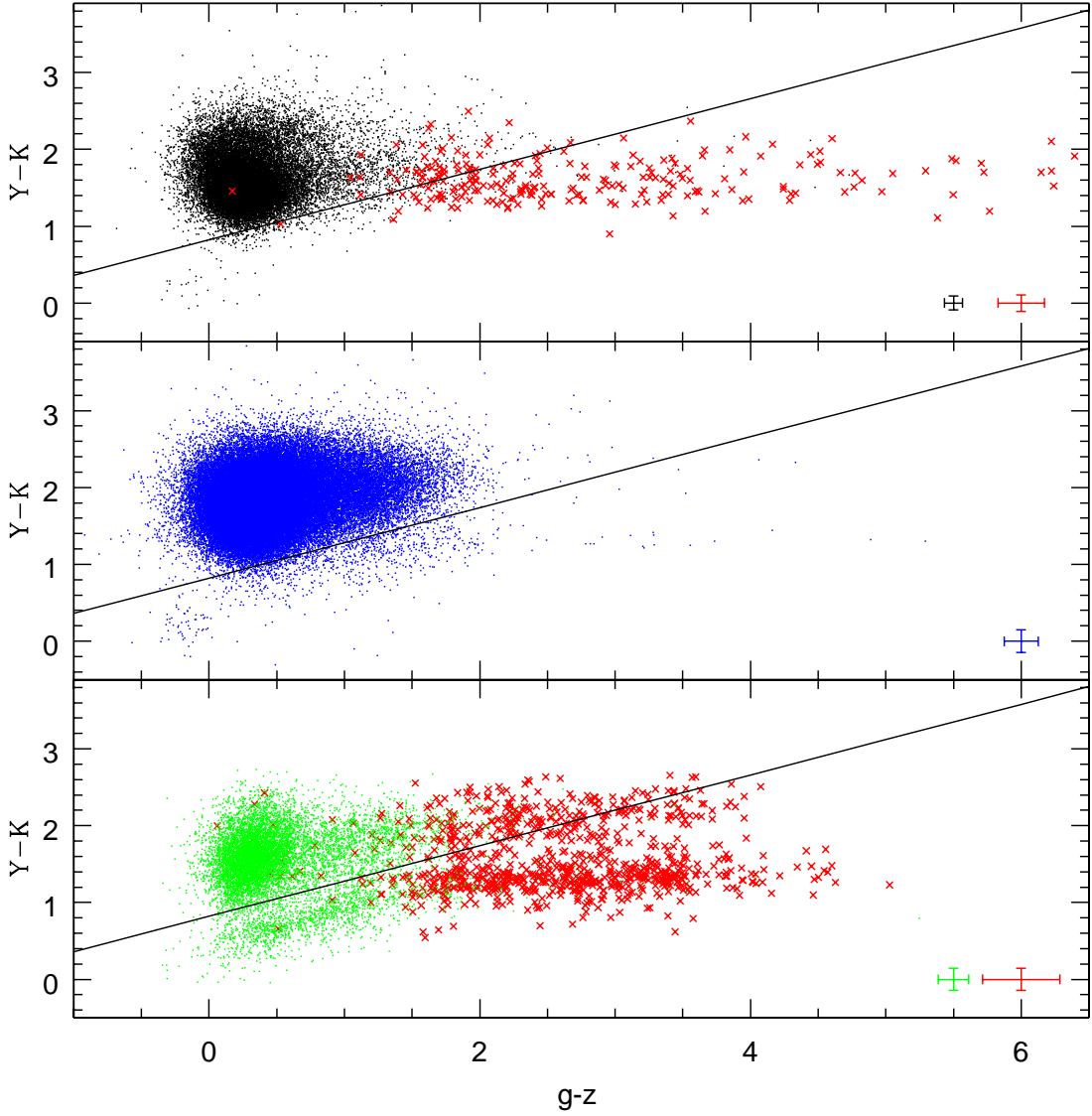}
\caption{Upper panel: The distribution of 24,878 known quasars in R09 in the Y-K/g-z color-color diagram. 
Black dots represent $z<4$ quasars and red crosses represent $z>4$ quasars. Middle panel: The distribution of  61,489 unidentified quasar
candidates in R09 with photometric redshift $z_{ph}<2.2$ in the Y-K/g-z color-color diagram. Lower panel: The distribution of  10687 unidentified quasar
candidates in R09 with photometric redshift $2.2<z_{ph}<4$ (green dots) and and 835 unidentified quasar
candidates in R09 with photometric redshift $z_{ph}>4$ (red crosses) in the Y-K/g-z color-color diagram. The error bars in the lower-right part of each panel denote the typical color uncertanty of quasars at different redshifts.} 
\end{figure}

\begin{figure}
\plotone{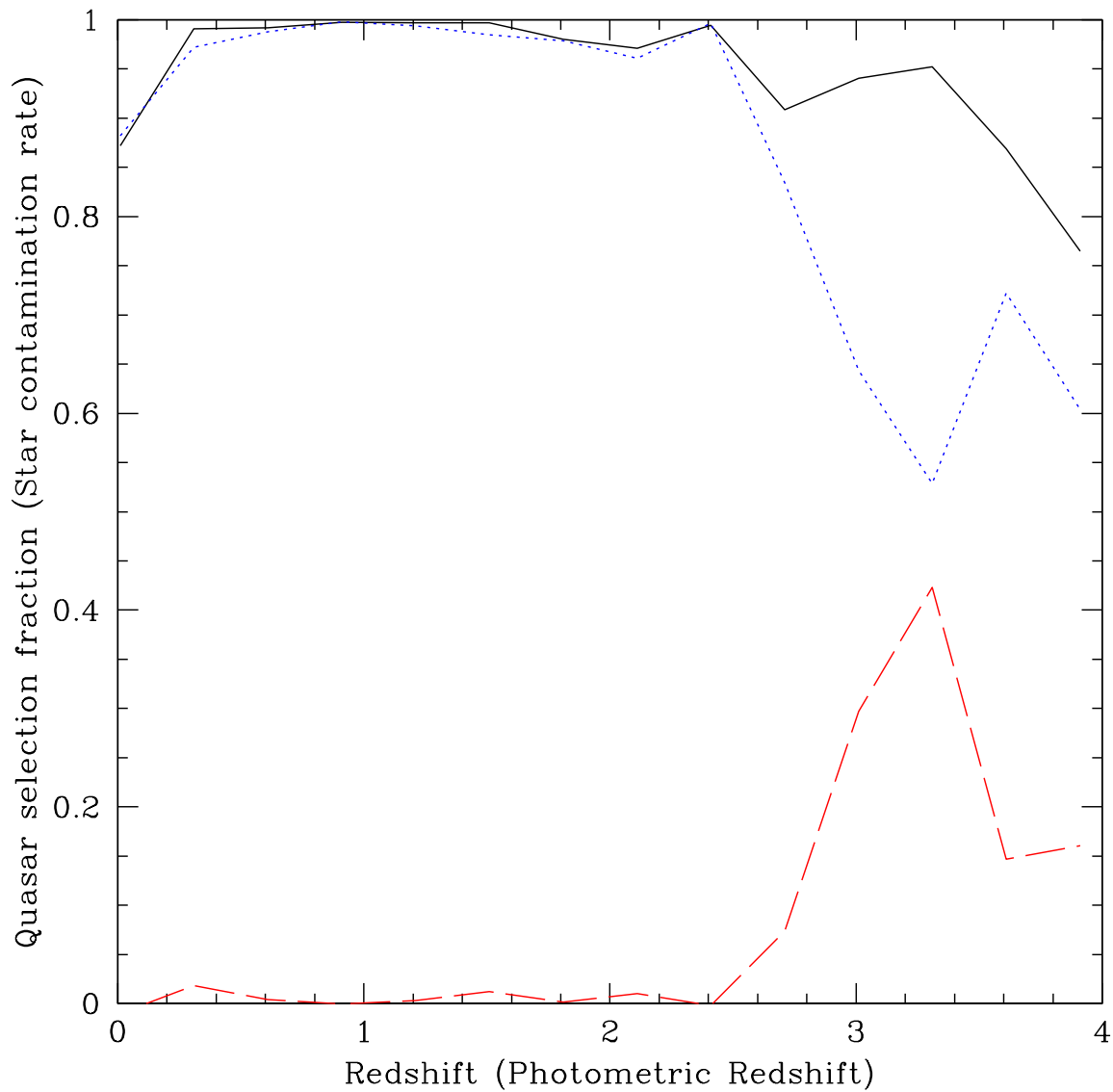}
\caption{The black solid line denotes the redshift dependence of the fraction of 24648 known $z<4$ quasars in R09 selected by the Y-K/g-z criterion. The blue dotted line denotes the fraction of 72,176 unidentified quasar candidates in R09 with photometric redshift $z_{ph}<4$ selected by
the Y-K/g-z criterion as a function of photometric redshifts. The red dotted line denotes the possible star contamination rates of these unidentified quasar candidates in R09 at different photometric redshifts.} 
\end{figure}

\begin{figure}
\plotone{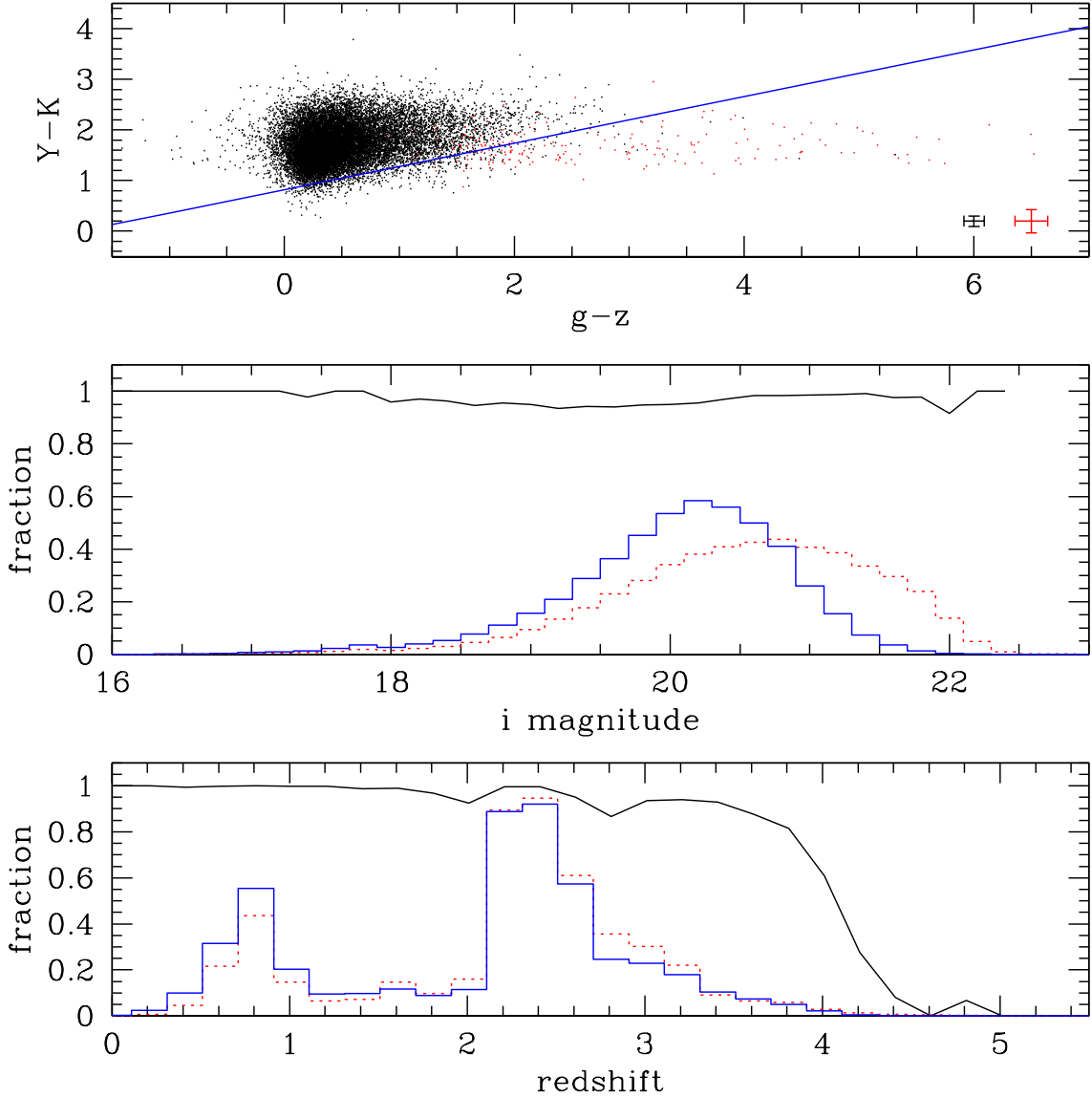}
\caption{Upper panel: The distributions of 17,999 DR9-UKIDSS quasars in the Y-K/g-z color-color
diagram. Quasars with redshift smaller (higher) than 4 are denoted as black (red) points. The black (red) error bars in the lower-right part denote the typical color uncertanty of quasars with redshifts smaller (higher) than
4. Blue line represents our proposed Y-K/g-z selection criterion for $z<4$ quasars.
Middle panel: Black line represents the magnitude dependence of the selection fraction of 17,999 DR9-UKIDSS quasars by the Y-K/g-z criterion. The blue and red histograms show the normalized magnitude distribution (fraction between the number of quasars in each magnitude bin and the total number) for 17,308 quasars selected by the Y-K/g-z criterion and  for 87,822 DR9 quasars, respectively. 
Lower panel: Black line represents the redshift dependence of the selection fraction of 17,999 DR9-UKIDSS quasars by our Y-K/g-z criterion. The blue and red histograms show the normalized redshift distribution (fraction between the number of quasars in each redshift bin and the total number) for 17,308 quasars selected by the Y-K/g-z criterion and  for 87,822 DR9 quasars, respectively. For clarity all histograms in the middle and lower panels are magnified by a factor of 5. }
\end{figure}

\begin{figure}
\plotone{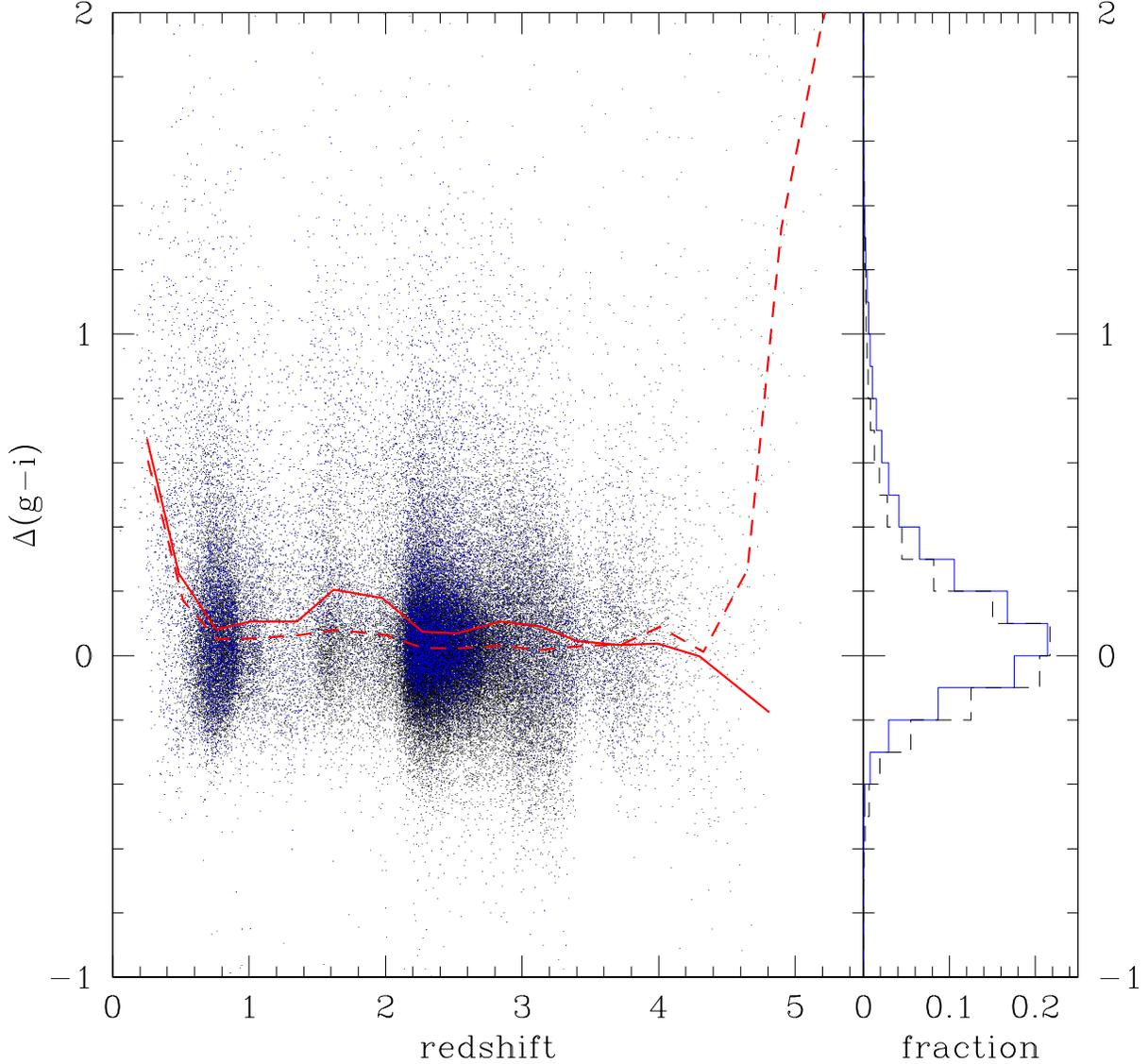}
\caption{The distributions of $\Delta(g-i)$ versus redshift for 17,308 quasars selected by the Y-K/g-z criterion (blue points) and  for 87,822 SDSS-III/DR9 quasars (black points). The red solid and dashed lines
show the distribution of median  $\Delta(g-i)$ values  in each redshift bin for these two quasar samples. The blue solid and black dashed histograms shows the distributions of  normalized $\Delta(g-i)$ (fraction between the number of quasars in each $\Delta(g-i)$ bin and the total number of quasars) in the whole range of redshift of  these two samples. The quasar sample selected by the Y-K/g-z criterion is slightly redder than the  SDSS-III/DR9 quasar sample at $z<4$.}
\end{figure}

\begin{figure}
\plotone{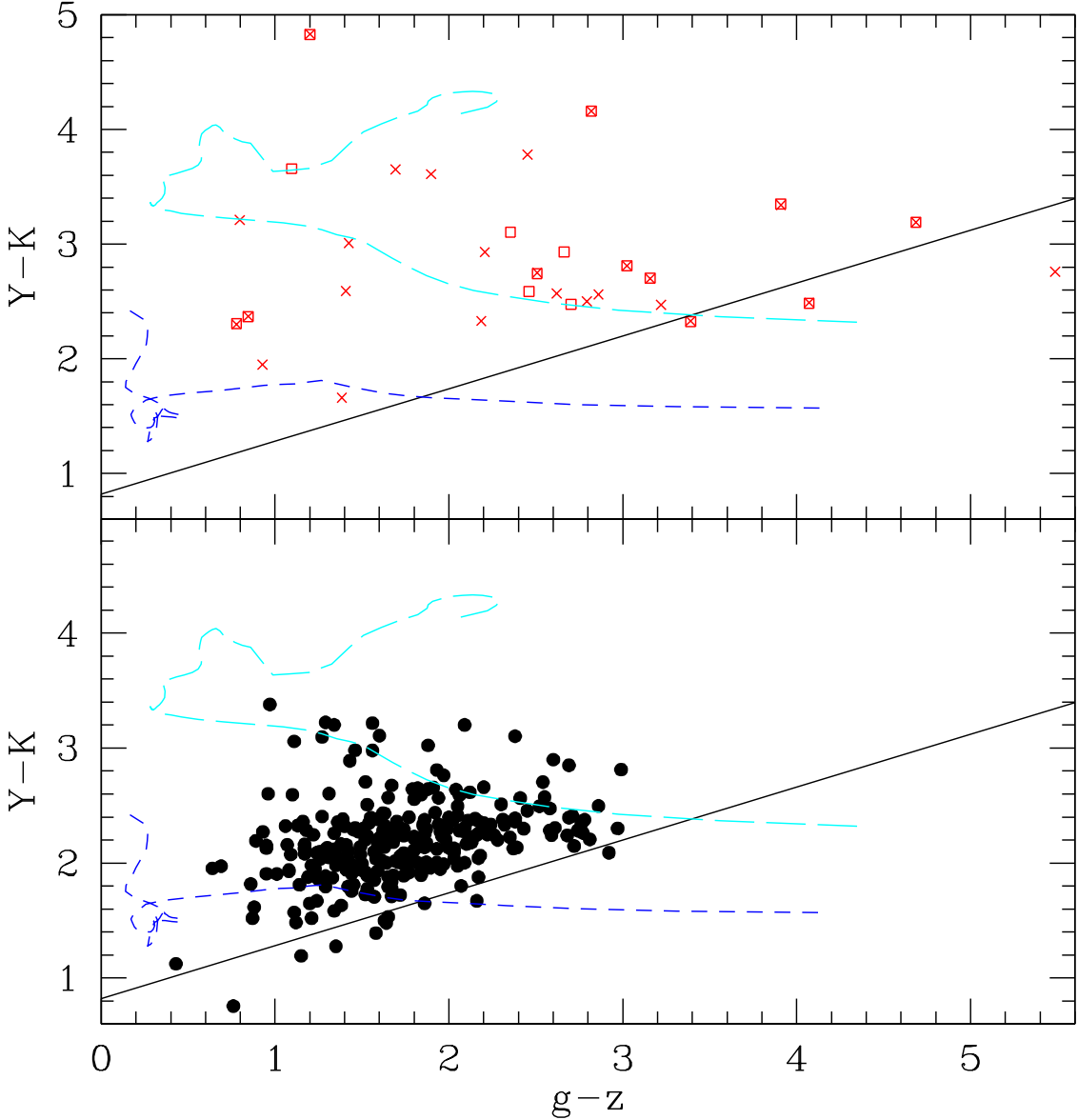}
\caption{Upper panel: Red quasars with UKIDSS data in the Y-K/g-z diagram. Red crosses and squares denote red quasars from Glikman et al. (2012) and Urrutia e al. (2009) respectively. There are 11 common quasars in these two samples. Lower panel: Optically selected type II quasars with UKIDSS data in the Y-K/g-z diagram. Filled circles denote type II quasars from Reyes et al. (2010). In both panel, black lines denote the Y-K/g-z selection criterion. Blue and cyan dashed lines represent the predicted colors of type I quasars and type II quasars, respectively,  at different redshift (up to $z=4.3$ to the right) using the templates from Polletta et al. (2007).}
\end{figure}

\end{document}